\documentstyle[twocolumn,aps]{revtex}
\begin{document}
\title{\Large{\bf Lattice Distortions and Charge Carriers in Cuprates}}
\author{{\bf Lev P. Gor'kov$^{1,2}$}}
\address{$^1$NHMFL, Florida State University, Tallahassee, FL 32310, USA}
\address{$^2$Landau Institute for Theoretical Physics, Moscow, Russia}
\maketitle
\pagestyle{empty}

\vspace{.25in}

\begin{abstract}

\noindent A phenomenological model with itinerant bands and local states
trapped by the lattice on the Cu-sites, is discussed to describe global
features of cuprates.  Relative energy positions of localized and
itinerant states being tuned (thermodynamically or by doping), the
system must undergo 1st order Mott metal-insulator transition.
Decreasing the local level (from the metallic end of a stoichiometric
compound), charge separation instability occurs first before the Mott
transition.  Crossing and hybridization between local (flat) and
itinerant bands introduce a structure in density of states which may
account for ``pseudogap'' features in cuprates.  Model results in
polaronic lattice effects and is rich enough to serve as a
phenomenology of cuprates.
\end{abstract}

\vspace{.25in}

\noindent{\bf 1. INTRODUCTION}

\vspace{.25in}

More than 10 years of intense studies of HTS after the seminal discovery
by G. Bednorz and A. M\"{u}ller have led to unprecedented amount of
experimental results.  Numerous theoretical views have been developed.
In that follows, an attempt is undertaken to unify physical trends and
theoretical ideas in a sort of  ``coarse-grained'' description, mixing
together microscopical statements and a phenomelogical interpretation.

Vast efforts are currently concentrated on interpretation of the
pseudogap phenomenon seen in the so-called underdoped regime in numerous
NMR-, transport-, and magnetization-experiments, and, more recently, in
the ARPES studies.

As for the ``overdoped'' regime, a consensus is currently growing [1,2] that
the farther one goes away from the ``optimal'' point, the clearer is the
trend to a restoration of the Fermi liquid (FL) state, judging at least,
by superconducting properties.

Let us take this view seriously.  In the stoichiometric compound
YBa$_2$Cu$_4$O$_8$, for instance, the underdoped-to-overdoped-transition
may be accessed by external pressure.  The 123 compound with O$_7$,
although it suffers from some structural instability [3], would be
another example of the material with the stoichiometric composition.  If
the FL views were correct, the latter, in accordance to the Luttinger's
counting theorem, should be a compensated metal, while the former- a
metal with the excess of one electron per unit cell.  The most popular
compound, Bi$_2$Sr$_2$CaCu$_2$O$_{8+\delta}$, viewed in the FL-terms at
$\delta\equiv 0$, would provide another case of compensated metal.
These differences experimentally seem to be of no large importance,
henceforth, we conclude that bands or pieces of the Fermi surface (FS)
involved into phenomena discussed below, in all cases are of the same
origin.   The usual assumption is that
the conduction bands in the CuO$_2$-plane are of the utmost importance,
but this detail will not be essential below.

Note that there is no signature of any presence of localized spins in
these (stoichiometric) materials.  On the other hand, in case of
La$_2$CuO$_4$ or YBa$_2$CuO$_6$ one meets the limit of the Mott
insulator.  It is also known that the doping process e.g., in
La$_{2-x}$Sr$_x$CuO$_4$, above $x>0.08$ results in the metallic regime,
corresponding to a large FS (with the area proportional to $1-x$, not to
x) with rather ordinary density of states (DOS).  We conclude, hence,
that doping by Sr realizes the Mott insulator-to-metal transition.  We
expect that such the transition may occur even in stoichiometric case,
provided a thermodynamical variable (such as pressure) could be found to
tune the transition.

Mott himself gave theoretical arguments that such a transition for a
perfect lattice must be of the $I^{st}$ order.  The symmetry arguments
allow to liken it to sort of a liquid-to-solid transition.  

Driving the Mott transition by a doping process is not a thermodynamical
procedure: dopants change the chemical composition and introduce
disorder, in addition, which would mask physical phase transition.

There are currently numerous theoretical models under discussion (for a
review see [4]).  Many
of them accept the lattice's involvement into properties of cuprates as
a key ingredient, by using the polaron-bipolaron's concepts.  Equally,
there are theories exclusively based on a single band model and strong
electron-electron correlations.  There are also  attempts to mix
different views together.  Remarkably enough, numerous experimental
trends quite often can be equally well explained from apparently
opposite points of view.  It seems to the author that main difficulties
with each of these theories lie in pursuing the underlying models to the
extreme, where new qualitative features begin to be important.  It
seems that a more global view  based on mixing microscopic
features with a phenomenological approach could be helpful.  

Such an attempt is suggested below.  Note, however, that we assume
importance of the lattice effects {\it to be out of the question}.
Thus, the isotopic effect {\it is seen} in
cuprates, although, and it is especially remarkable, on the so-called
``underdoped'' side.  There is convincing experimental evidence for
presence of lattice effects both in the normal and superconducting states
(for the excellent summary see [5]).  It is our hope to show that other
theoretical views may stem out of these notions.

\vspace{.25in}

\noindent{\bf 2. MODEL}

\vspace{.25in}

We distract ourself from the disorder introduced by
dopants, and address the metal-to-insulator-transition issue as if it
could have occured in a stoichiometric material.  The presence of
localized Cu$^{2+}$ states (spins) is considered as a hallmark of the
insulator phase.  To describe the Mott transition, we consider
the model first suggested in [6].  This model has been an
extension of the intuitive notions which have led Bednorz and M\"{u}ller
to their discovery.  It is a ``two component'' model.  It has been shown
in [6] for the first time that the Mott phase transition must be preceded
by the charge phase separation instability, which plays a key role
below.

In the model there are two types of electronic (hole) sites.  The
localized ones are basically associated with the each Cu-sites (in the
plane).  If a local level with an energy $\varepsilon_f$ was occupied by
one hole $(d^{9};~Cu^{2+})$, being well below any ``empty conduction''
band, this state would correspond to the insulating end.  On the other
end, in the ideally metallic state itinerant bands are properly filled
up.  There are no holes on the levels $\varepsilon_f$.  Level
$\varepsilon_f$ being tuned down, holes may begin to occupy the local
states by thermal activation, or, if $\varepsilon_f$ goes below the
initial chemical potential, a non-zero concentration of occupied local
centers appears even at $T=0$.  The physical idea for charge phase
separation was that an occupied center {\it inevitably} produces local
lattice deformations.  Two levels may relieve elastic energy by coming
closer to each other.  Hence, the tendency to form clusters (the
charge-phase separation) would be an immediate consequence of such an
attraction which, however,  is limited by the electroneutrality condition.
Therefore clusters may exist in a virtual form, resulting in a ``foggy''
state as described in [6].  (If the electroneutrality can be achieved by
redistribution of mobile oxygen ions, the actual phase transition may
occur as it, in fact, takes place in the oxygen rich
La$_2$CuO$_{4+\delta}$).

The charge phase separation instability at small enough
$|\varepsilon_f|$ is not to be confused with the Mott transition.
Although the lattice  energy brings local centers together, the Coulomb
repulsion still keeps them apart.  The clusters mentioned above are
rather loose formation (on the scale of two-four interatomic distances,
as judged from the miscibility gap value $\delta =0.06$ for the phase
separation in the oxygen rich La$_2$CuO$_{4+\delta}$).  The energy
barrier related to the Mott transition, corresponds to the energy cost
of bringing local centers close together.  For that the level
$\varepsilon_f$ must sink deep down into the filled-up band.

The concentration of populated centers, $n_f$, is determined by the
obvious relation:
\begin{eqnarray}
n_f +n_h =n_0
\end{eqnarray}

\noindent where $n_0$ is the total number of holes per unit cell.  There is no
double occupancy  on the local site:
\begin{eqnarray}
n_f=2e^{-(\varepsilon_f-\mu)/T}\cdot \left\{
1+2e^{-(\varepsilon_f-\mu)/T}\right\}^{-1} 
\eqnum{1'}
\end{eqnarray}

\noindent The number of the hole-occupied local centers is exponentially
small at $\Delta=\varepsilon_f-\mu >0$ and $T\ll\Delta$.  On the other
hand, it is easy to see from $(1,1')$ that at $ T\gg |\Delta |$:
\begin{eqnarray}
n_f\propto \left(T/\mu_0\right)\ln\left( \mu_0/T\right)
\nonumber
\end{eqnarray}
 
\noindent Scattering of the band holes by these centers is one of the
sources of resistivity at high enough temperatures.

\vspace{.25in}

\noindent{\bf 3. HYBRIDIZATION AND FERMI SURFACE}

\vspace{.25in}

Leaving aside for awhile center's tendency to cluster (which is due to
lattice effects and is to be characterized by another energy
scale), one may consider the problem of local centers in frameworks of
the periodic Anderson  model, if a hybridization between localized and
itinerant states is present.  One should be cautious, however.  Our
assumption was that local levels if occupied by a hole, produce local
deformations and may be due to the latter itself, at least to some extent.  In
other words, a polaronic aspect is already here.  On the other hand, the
thermodynamical equilibrium in eq.(1) establishes itself through exchange by
holes between itinerant and local states.  Therefore, the Anderson model
is applicable to the extent that hopping, $V$, is large enough to
neglect a slow motion of the lattice degrees of freedom.  Hence, the assumption
\begin{eqnarray}
V>\omega_0
\end{eqnarray}

\noindent where $\omega_0$ is a characteristics phonon frequency scale.

The periodic Anderson model has no exact solution yet.  However, the
basic physics, sufficient for the discussion here, becomes
clear from the mean field solution [7,8].  Namely, let $\varepsilon ({\bf
k})$ be the initial band spectrum, and write the hybridization Hamiltonian in
the form:
\begin{eqnarray}
\sum_{\bf k} \tilde{V}_{\bf k}(\hat{c}^+_{\bf k}\hat{d}_{\bf k} +
\hat{d}^+_{\bf k}\hat{c}_{\bf k})
\end{eqnarray}

\noindent where the ``bare'' hybridization comes together with the
factor $(1-\bar{n}_f)^{1/2}$ accounting for single occupancy of the
$\varepsilon_f$-level: 
\begin{eqnarray}
\tilde{V}_{\bf k}=(1-\bar{n}_f)^{1/2}V_{\bf k}
\end{eqnarray}

\noindent ($\bar{n}_f$- the average occupancy of the
$\varepsilon_f$-level).  For the hybridized spectrum one obtains
\begin{eqnarray}
\varepsilon_{\pm} ({\bf k}) = \varepsilon_f+\frac{1}{2}(\varepsilon({\bf
k})-\varepsilon_f \pm \nonumber \\
\sqrt{(\varepsilon({\bf
k})-\varepsilon_f)^2+4|\tilde{V}_{\bf k}|^2}~ ) 
\end{eqnarray}

\noindent The chemical potential lies inside the $\varepsilon_-({\bf
k})$-''conduction'' band $(T=0)$ preserving the total number of
carriers.  Hybridization between one flat $(\varepsilon_f)$ and one
dispersive band results in only minor changes of the FS at
$\varepsilon_f>\mu_0$ ($\mu_0$-the bare chemical potential), while the
``flattening'' of the $\varepsilon_-({\bf k})$ at $\varepsilon_f<\mu_0$
increases rapidly when $\varepsilon_f$ goes deeply enough under $\mu_0$.
The new Fermi momenta, ${\bf k}_F$, may then lie far apart from the
``crossing'' points, ${\bf k}_0$, defined by $\varepsilon({\bf
k}_0)=\varepsilon_f$.  Recall that in the mean field approximation the
``new'' FS is still determined from conservation of the
total holes number.

The Green function for the conduction band is of the form [8]:
\begin{eqnarray}
G_c({\bf k}, \omega) =\frac{v_{\bf k}^2}{\omega-\varepsilon_{-}({\bf k})}
+\frac{u_{\bf k}^2}{\omega-\varepsilon_+({\bf k})}
\end{eqnarray}

\noindent where the factors, $u_{\bf k};~v_{\bf k}$, are, as usual:
\begin{eqnarray}
v^2_{\bf k}=\frac{1}{2} \left\{ 1- (\varepsilon({\bf
k})-\varepsilon_f)/E({\bf k})\right\} \nonumber \\
u^2_{\bf k}=\frac{1}{2}
\left\{ 1+ (\varepsilon({\bf k})-\varepsilon_f)/ E({\bf k})\right\}
\end{eqnarray}

\noindent (here $E({\bf k})$ stands for the square root in (5)). 

Allusions to recent low temperature ARPES-results [1,2,9] are self evident.
First of all, factor $v^2_{\bf k}$ smears away the sharpness of $\mbox{Im}
G_c({\bf k}, \omega)$ at the Fermi surface on distances
$|\varepsilon({\bf k})-\varepsilon_f|\sim |\tilde{V}_k|$.

Secondly, one may have chosen an anisotropic $V_{\bf k}$, say:
\begin{eqnarray}
V_{\bf k}=V_0(\cos k_xa-\cos k_ya)
\end{eqnarray}

\noindent Under assumption (8) most profound effects are expected to
take place in a vicinity of the point $(0,\pi)$ of the Brillouin zone.
Changes in the shape of FS certainly depend on the ``bare'' spectrum,
$\varepsilon({\bf k})$, near $(0,\pi)$.  A more detailed discussion will
be published elsewhere.  Here we content ourselves with the remark that
the decrease of the residue, $v_{\bf k}^2$, in eq. (6) at approaching
${\bf k}_f$ across the intersection point, ${\bf k}_0\left(\varepsilon({\bf
k}_0)=\varepsilon_f\right)$, qualitatively accounts for smearing of the
normal state excitation peak and appearence of the ``leading edge''
feature seen in the ARPES-experiments [1,2] for the 2212-compound along
the X-M-direction.  If the above is indeed related to
phenomena seen in [1,2,9], one may expect $V$ of order of 10-40 meV, depending
on the material.

Further ``heaviness'' (flatness of the $\varepsilon_{-}({\bf k})$-band at
$k_F$) would proceed with the further sinking of the local level into
the broad band resulting in the increase of the density of states and
other Kondo-like features well known from the Heavy Fermions (HF)
physics.  It suppresses charge fluctuations by reducing $\tilde{V}_{\bf k}$
and breaking down (2).  The Born approximation ceases to be valid, and
non-adiabaticity results in the polaronic factors of the form
\begin{eqnarray}
\exp (-W/\omega_0)
\end{eqnarray}

\noindent (with $W$ being on the bandwidth scale) weakening the rate of transitions between localized and itinerant
states.  The physics of mixed valency would not work anymore because
transitions between occupied and unoccupied states for the local center
now involve inelastic lattice processes.  The structure of the
conduction bands of eq. (5) around ${\bf k}_F$ gets ``saturated'' at
$\tilde{V}_{\bf k}\sim\omega_0$.  The flat incoherent portions
correspond to polarons which may be thermally activated to form an
``occupied site'' in a vicinity of the Cu-ions.  Quantum-mechanically,
their evolution is slow (i.e., is on the $\omega_0$-scale).

\vspace{.25in}

\noindent {\bf 4. CLUSTERS}

\vspace{.25in}

As mentioned above, there is a tendency for the occupied local centers
to cluster gaining in the elastic (lattice) energy.  On the other hand,
the Coulomb energy prevents occupation of local centers on adjacent
Cu-sites.  Therefore, such a cluster being a nucleation center of a new
phase, is actually a rather complicated object which may live on a time
scale even longer than $\sim\hbar /\omega_0$.  Lattice and charge
degrees of freedom are strongly coupled in it.  However, it is important
to emphasize that such an object cannot reach a macroscopic scale
without causing a 
violation of the electroneutrality in underlying lattice [6,10].  Here lies the
major difference with notions regarding formation of self-trapped
polarons or bipolarons.  The cluster may be comprised of a large number
of occupied local centers but still has a {\it finite} life time for its
existence.

There is no shortage in theoretical studies of ``negative-$U$'' centers,
or superconductivity based on mixed valence or bipolaronic mechanisms [5,11].
In our view, they all suffer from too specific assumptions.  In
particular, they do not include the interplay between
lattice and the physics of the periodic Anderson model.

It seems that ideas used for treatment of nuclear reactions in heavy
nuclei may provide us with a helpful phenomenological approach.  The
Bohr's concept of a {\it compound (composite) nucleus} assumes that a
scattering or a nuclear reaction, ($i$), passes through the initial stage 
at which an incident particle colliding with a nucleus, is first trapped
by it, exchanging by energy with other particles inside.  At the final
stage $(f)$ the {\it compound nucleus} emits a few over particle, or the
same particle.   For a specific channel $(i, f)$ the
partial amplitudes, $f_{if}(\omega)$, may be written in the form [12]:
\begin{eqnarray}
f_{if}(\omega)=\frac{\Gamma(\gamma_{if}/k_F)}{\omega -E_0+i\Gamma}
\end{eqnarray}

\noindent where $\gamma_{if}$ is a matrix element for a specific
``reaction'', while $\Gamma$ represents the total width of the resonance
at some $E_0$.  The contribution from scattering on clusters into  the
imaginary part of the self-energy is then
\begin{eqnarray}
\sum (\omega ,{\bf k})\propto\sum_f\int d\omega'd{\bf k}'|f_{if}(\omega
-\omega')|^2 \cdot \nonumber \\ \mbox{Im} G(\omega',{\bf k'})
\end{eqnarray}

\noindent   Summation (over $f$) in
(11) corresponds to all possible processes at interaction of an
electron, ${\bf k}$, with a cluster.  At least at high temperatures the
main contribution into the e-e-scattering would be due to inelastic
processes.

\vspace{.25in}

\noindent {\bf 5. SUPERCONDUCTIVITY CHANNEL}

\vspace{.25in}

Let us discuss in some detials the Cooper channel. Note that, according
to (3), two electrons $({\bf k}, -{\bf k})$ join a cluster through
hybridization matrix element, $\tilde{V}$.  Generally, not only two (as
in bipolaron approach) but any number of particles may be virtually
trapped into the cluster, including odd numbers.

Scattering in the Cooper channel can be singled out with respect to
others due to the Cooper phenomena itself which is the basis of the
BCS-theory.  Another possibility, in Eq. (10) the resonance amplitude
prevails in the two particle channel.  The latter sounds rather similar to
current ideas regarding ``pre-paired'' states or to the bipolaronic
physics.  (Virtual two-particle bound states have first been introduced
in [13]).  Note, however,  that unlike the concept of stable bipolarons, these
states may be only temporarily occupied, according to (10).

If, as we assume, the ``attraction'' in the Cooper channel is due to the
tendency to cluster, it is worth mentioning that the mere assumption of
anisotropic hybridization, i.e. $V_{\bf k}$ from eq. (8), is already
enough to produce symmetry of, at least $|\Delta({\bf k})|$, similar to
that one of the pseudogap.

In the classic BCS-theory the Cooper channel is singled out by the
logarithmic singularity, $g_{eff}\ln \left(\bar{\omega}/T_c\right)\sim
1$, where $\bar{\omega}$ is a cut-off and $g_{eff}$ is the effective
interaction strength.  In frameworks of the above scheme, in the
metallic regime $\varepsilon_f$ is first well above the chemical
potential (and so is $E_0$ in (10)).  With $\varepsilon_f$ going down,
$f_{{\bf k}, -{\bf k}}$ in (10) $(\propto g_{eff})$ increases and so
does $T_c$.  In the BCS-like expression each of $\varepsilon_f, E_0,$
and even $\Gamma$ from (10) may play the role of a cut-off,
$\bar{\omega}$.  It is more difficult to interpret the superconductivity
in terms of a BCS-like picture on the ``underdoped'' side.  The decrease
of $T_c$ may be ascribed to the disruption of communication between
itinerant electrons and increased clusters of the ``new phase'' (the
polaronic factors in eq. (9)).  Experiment shows the most pronounced
isotope effect namely in the ``underdoped'' regime, which qualitatively
agrees with this speculative idea.

\vspace{.25in}

\noindent {\bf 6. STRIPES}

\vspace{.25in}

The phase separation has been considered so far as a virtual process in
a stoichiometric material driven by motion of the localized level,
$\varepsilon_f$, downward into the conduction band.  An amount of the
``foggy'' phase, however, may result in sort of self-organized
intermittent structure, which currently became a very active topic
[14,15].  Although lattice effects may easily produce even a static
super-structure, including stripes, there are no means to address the
issue on a quantitative basis, especially, if disorder is to be taken
into account.  The most direct evidence favoring a mechanism of stripes
formation, is the physical phase separation which occurs in the oxygen
enriched La$_2$CuO$_{4+\delta}$.  From $\delta\simeq 0.06$ one may
estimate effective distance between single occupied Cu-sites as two-four
lattice distances.  Superstructure for the oxygen-rich phase has been
seen, indeed, in neutron experiments both along the $c-$ and
in-plane-directions [16,17].

\vspace{.25in}

\noindent {\bf 7. CONCLUSIONS}

\vspace{.25in}

We suggest a unifying view for different regimes in cuprates on the
metallic side of the Mott transition in terms of the relative position
of a localized level, $|\varepsilon_f|$, assumed to reside mostly on the
Cu-sites, with respect to the chemical potential in an itinerant band.

This view has already  led to the conclusion [6] that a charge phase separation
is to precede the Mott transition due to the obvious fact that lattice
energy of a few such local sites would  diminish by forming
clusters.

The size of clusters being limited by the electroneutrality provision,
is probably rather small.  As the result, the clusters may form a
``foggy'' state, where quantum mechanical fluctuations play an important
role.  Among options to minimize the Coulomb effects by
forming static or quasistatic periodic structure, stripes are one of
possibilites.  The charge
phase separation manifests itself in oxygen-rich La$_2$CuO$_4$.  

Intersection of the flat ``band'' formed by local levels, with an
itinerant band results in peculiarities of the density of states,
reminiscent of the ``pseudogap'' phenomena, if an appropriate
anisotropic hybridization is present.

Although there are many polaronic features implicitly involved into the
phenomena, polarons or bipolarons themselves, as stable charge carriers,
are not necessary for the main physics.

Mechanism of superconductivity in which pairing goes through the virtual
trapping of the Cooper electrons by clusters, was discussed in a
phenomenological manner.

For different regimes the suggested view may be reconciled with a number
of microscopic models.

\vspace{.25in}

\noindent {\bf ACKNOWLEDGMENTS}

\vspace{.25in}

The author benefited from helpful discussions with I. Bozovic, A.
Chubukov, J. Dow, and J.R. Schrieffer.

This work was supported by the NHMFL through NSF cooperative agreement
No. DMR-9016241 and the State of Florida.

\vspace{.25in}

\noindent {\bf REFERENCES}

\vspace{.25in}

\noindent [1] A.G. Loeser {\it et al., Science} {\bf 273}, 325 (1996).

\noindent [2] H. Ding {\it et al., Nature} (London) {\bf 382}, 51 \\
\indent $~~$(1996).

\noindent [3] E. Kaldis {\it et al., Phys. Rev. Lett.}, {\bf 79}, 4894
\\ \indent $~~$(1997).

\noindent [4] R. Micnas and S. Robaszkiewicz, in {\it ``High-}\\ \indent
{\it $~~T_c$
Superconductivity 1996: Ten Years} \\ \indent $~~${\it After the
Discovery,''} E. Kaldis {\it 
et al.}, eds. \\ \indent $~~$NATO ASI Series (Kluwer Academic
\\ \indent $~~$Publishers, London, 
1997), p. 31.

\noindent [5] D. Michailovic and K.A. M\"{u}ller, {\it ibid.} p. 243; \\ \indent $~~$D.
Michailovic, p. 257.

\noindent [6] L.P. Gor'kov and A.V. Sokol, {\it JETP Lett.}, {\bf 46},
\\ \indent $~~$420 (1987).

\noindent [7] P. Coleman, {\it Phys. Rev.} {\bf B29}, 3035 (1984).

\noindent [8] A.J. Millis and P.A. Lee, {\it Phys. Rev.} {\bf B35}, \\
\indent $~~$3394 (1987).

\noindent [9] N.L. Saini {\it et al., Phys. Rev. Lett.} {\bf 79}, 3467
\\ \indent $~~$(1997).

\noindent [10] V.J. Emery, S.A. Kivelson, and H.-Q. Lin, \\ \indent $~~~${\it
Phys. Rev. 
Lett.} {\bf 64}, 475 (1990); S.A. \\ \indent $~~~$Kivelson, V.J. Emery,
and H.-Q. Lin, 
{\it Phys.}\\ \indent $~~~${\it  Rev.} {\bf B42}, 6523 (1990); V.J. Emery
and S.A. \\ \indent $~~~$Kivelson, 
{\it Physica} {\bf C209}, 597 (1993).

\noindent [11] B. Brandow, {\it Phys. Rep.} {\bf 296}, 1 (1998).

\noindent [12] L.D. Landau and E.M. Lifshitz, {\it Quantum}\\ \indent
$~~~${\it  Mechanics:
non-relativistic theory} (Perga-\\ \indent $~~~$mon Press, New York, 1977).

\noindent [13] S.P. Ionov, {\it Sov. Phys.-- Izvestia Akad.}\\ \indent
$~~~${\it  Nauk, Seria
Fiz.} {\bf 49}, 90 (1985); G.M. \\ \indent $~~~$Eliashberg, {\it JETP Lett. Suppl.} {\bf
46}, 5181 \\ \indent $~~~$(1987).

\noindent [14] J. Tranquada {\it et al., Nature} (London) {\bf 375}, \\
\indent $~~~$561
(1995); {\it Phys. Rev.} {\bf B54}, 7489 (1996); \\ \indent $~~~$J.M. Tranquada, J.D.
Axe, N. Ichikawa, \\ \indent $~~~$A.R. Moodenbaugh, Y. Nakamura, and S.
\\ \indent $~~~$Uchida, {\it
Phys. Rev. Lett.} {\bf 78}, 338 (1997).

\noindent [15] N.L. Saini {\it et al., Phys. Rev.} {\bf B57}, R11101
\\ \indent $~~~$(1998).

\noindent [16] B.O. Wells {\it et al., Science} {\bf 277}, 1067 (1997).

\noindent [17] X. Xiong {\it et al., Phys. Rev. Lett.} {\bf 76}, 2997
\\ \indent $~~~$(1996).

\end{document}